\documentclass{article}

\usepackage{cite}
\usepackage{amsmath,amssymb,amsfonts}
\usepackage{graphicx}
\usepackage{mlspconf}
\usepackage{textcomp}
\usepackage{xcolor}
\usepackage{caption}
\usepackage{algorithm}
\usepackage[noend]{algpseudocode}
\usepackage{scalerel}
\usepackage{subcaption}
\usepackage[nolist]{acronym}

\usepackage{tikz, circuitikz}
\usetikzlibrary{arrows,decorations.pathmorphing,backgrounds,positioning,fit,petri,quotes}
\usepackage{import}

\usetikzlibrary{svg.path}
\definecolor{orcidlogocol}{HTML}{A6CE39}
\tikzset{
  orcidlogo/.pic={
    \fill[orcidlogocol] svg{M256,128c0,70.7-57.3,128-128,128C57.3,256,0,198.7,0,128C0,57.3,57.3,0,128,0C198.7,0,256,57.3,256,128z};
    \fill[white] svg{M86.3,186.2H70.9V79.1h15.4v48.4V186.2z}
                 svg{M108.9,79.1h41.6c39.6,0,57,28.3,57,53.6c0,27.5-21.5,53.6-56.8,53.6h-41.8V79.1z M124.3,172.4h24.5c34.9,0,42.9-26.5,42.9-39.7c0-21.5-13.7-39.7-43.7-39.7h-23.7V172.4z}
                 svg{M88.7,56.8c0,5.5-4.5,10.1-10.1,10.1c-5.6,0-10.1-4.6-10.1-10.1c0-5.6,4.5-10.1,10.1-10.1C84.2,46.7,88.7,51.3,88.7,56.8z};
  }
}
\newcommand\orcidicon[1]{\href{https://orcid.org/#1}{\mbox{\scalerel*{
\begin{tikzpicture}[yscale=-1,transform shape]
\pic{orcidlogo};
\end{tikzpicture}
}{|}}}}

\usepackage{hyperref}
\hypersetup{
colorlinks=true, %
linktoc=all,     %
linkcolor=black,  %
}

\def\BibTeX{{\rm B\kern-.05em{\sc i\kern-.025em b}\kern-.08em
    T\kern-.1667em\lower.7ex\hbox{E}\kern-.125emX}}

\toappear{2025 IEEE International Workshop on Machine Learning for Signal Processing, Aug.\ 31-- Sep.\ 3, 2025, Istanbul, Turkey}

\title{Energy-Information Trade-Off\\in Self-Directed Channel Memristors}

\name{
Waleed El-Geresy \orcidicon{0000-0002-4016-6078}$^{\star}$
\qquad D\'aniel Hajt\'o \orcidicon{0000-0002-5815-2256}$^{\dagger}$
\qquad Gy\"orgy Cserey \orcidicon{0000-0002-6836-1502}$^{\dagger}$
\qquad Deniz G\"und\"uz \orcidicon{0000-0002-7725-395X}$^{\star}$
\thanks{This work was supported by the EPSRC DTP 2016-2017 (EP/N509486/1), DTP 2018-2019 (EP/R513052/1), SONATA (EP/W035960/1), and NKFIH (2019-2.1.7-ERA-NET-2021-00023) projects.}
}
\address{
$^{\star}$ Imperial College London \;
$^{\dagger}$  P\'azm\'any P\'eter Catholic University
}

\begin{document}

\maketitle

\begin{abstract}

    Understanding the nature of information storage on memristors is vital to enable their use in novel data storage and neuromorphic applications. One key consideration in information storage is the energy cost of storage and what impact the available energy has on the information capacity of the devices. In this paper, we propose and study an energy-information trade-off for a particular kind of memristive device - \ac{SDC} memristors. We perform experiments to model the energy required to set the devices into various states, as well as assessing the stability of these states over time. Based on these results, we employ a generative modelling approach, using a \ac{cGAN} to characterise the storage conditional distribution, allowing us to estimate energy-information curves for a range of storage delays, showing the graceful trade-off between energy consumed and the effective capacity of the devices.

\end{abstract}

\acresetall

\section{Introduction and Motivation}

Memristors are a type of passive electronic device typically defined by their capacity to switch between resistive states. Initially proposed by Leon Chua in 1971 to complete a ``periodic table" of passive electronic components \cite{chuaMemristorMissingCircuit1971}, their potential for describing neuronal dynamics was soon highlighted \cite{chuaMemristiveDevicesSystems1976}. They subsequently emerged as candidate neuromorphic computing elements \cite{el-geresyEventBasedSimulationStochastic2024}, for the storage and in-place processing of information to overcome the Von Neumann bottleneck \cite{vonneumannFirstDraftReport1993}. One of the key advantages of neuromorphic computing is energy-efficient information processing. It is therefore essential to develop our understanding of information storage on memristors and how energy available impacts the information capacity of the devices, which plays a key role in enabling their use for intelligent computation.

Previous research identified a correlation between the magnitude of current pulses used to program memristive devices and the resulting resistance level achieved \cite{ielminiSizeDependentRetentionTime2010}, suggesting a trade-off between power consumption and the final resistance state.
In \cite{el-geresyEnergyConstrainedInformationStorage2024}, an explicit logarithmic trade-off was identified between the final conductance of a device (state) and the energy of the programming pulses applied.
Different memristive states can exhibit varying degrees of reliability, with some states being more stable than others due to stochastic resistive drift \cite{el-geresyEventBasedSimulationStochastic2024, el-geresyDelayConditionedGenerative2024}. The instability of these states, coupled with the varying amounts of energy required to program the devices into different states, creates an implicit trade-off between the energy consumed during programming and the information reliably recoverable after different delays.
Previous research has attempted to measure the capacity of various memristive devices in the presence of different non-idealities. A graph-theoretic approach to characterizing the capacity of binary memristors in the presence of sneak path current noise was investigated in \cite{rumseyCapacityConsiderationsData2019}. The capacity of \ac{PCM} devices for storage was empirically estimated in \cite{engelCapacityOptimizationEmerging2015, engelOpportunitiesAnalogCoding2017} using the Blahut-Arimoto algorithm \cite{blahutComputationChannelCapacity1972}.
However, these approaches did not consider the energy cost of states.

In this paper, we present a study of the information-theoretic limits of information storage under energy constraints on memristive devices in the presence of delay-conditional resistive drift noise. We focus on \ac{SDC} memristors, which are one of the few types of memristors currently commercially available~\cite{campbellSelfdirectedChannelMemristor2017}. In Section~\ref{sec:energy_characterisation}, we use an iterative optimisation procedure to estimate an energy cost function that yields the energy cost of different device initial states. In order to enable us to model the effect of data storage under different delay conditions, in Section~\ref{sec:state_stability}, we use a \ac{cGAN} to model the delay conditional storage distribution. We combine these results in Section~\ref{sec:blahut}, to estimate the cost-constrained capacity for a variety of storage delays. To the best of our knowledge, this is the first study of its kind to explore the explicit trade-off between energy and information capacity in memristors.

\section{Problem Formulation}
\label{sec:problem_formulation}

Our goal is to estimate a trade-off between the capacity of a memristor for information storage and the energy available to us to program the device. To this end, we must understand the states of the device, which can be used to store information.

\subsection{Memristive State and State Instability}

In \cite{hajtoStateCharacterisationSelfDirected2025}, a proposed memristive state was introduced for Knowm W+\ac{SDC} memristors, accounting for the non-linear VI characteristic of the devices. The state is characterised by the relation between the instantaneous voltage, \(v(t)\), and current, \(i(r, v)\):

\begin{align}
    i(r, v(t)) &= r \cdot (G_m \cdot v(t) + I_d(v(t)))
    \label{eq:vi}\\
    I_d(v(t)) &= \alpha_1 \cdot (e^{\beta_1 \cdot v(t)} - 1) + \alpha_2 \cdot (1 - e^{-\beta_2 \cdot v(t)})
    \label{eq:schockley_modified}
\end{align}
where \(G_m\), \(\alpha_1\), \(\alpha_2\), \(\beta_1\), and \(\beta_2\) are parameters that characterise the VI relation for a particular device, and \(x\) is the state variable that determines the instantaneous VI relation. \(I_d\) represents the diode component of the total current. In this paper, we use the model parameters from \cite{hajtoStateCharacterisationSelfDirected2025}, shown in Table~\ref{tab:model_parameters}.

\begin{table}[!htbp]
    \centering
    \begin{tabular}{ccccc}
        \(\alpha_1\) & \(\alpha_2\) & \(\beta_1\) &  \(\beta_2\) & \(G_m\) \\
        \hline
        \(2.73 \cdot 10^{-3}\) & \(1.313 \cdot 10^{-7}\) &  \(13.92\) & \(2.327 \cdot 10^{-6}\) & \(4.207\) \\
        \hline
        \hline
    \end{tabular}
    \caption{Parameters for the \ac{SDC} memristor state model.}
    \label{tab:model_parameters}
\end{table}

However, the identified state is not stable and changes over time, even in the absence of a stimulating voltage. We can represent this mathematically as a conditional probability distribution, where the state after a delay \(D\), denoted \(\hat{R}\), is a random variable conditional on the initial state, denoted \(R\), and the delay, \(D\): \(\hat{R} \sim P(\hat{R}|R, D)\).
This formulation lends itself to a communication-theoretic treatment of the storage problem, where the conditional storage distribution can be seen as a delay-dependent communication channel, \(P(\hat{R}|R, D)\), with the initial state, \(R\), denoting the channel input. We can thus characterise the capacity of this storage channel, under given constraints.

\subsection{Energy-Constrained Capacity}

The capacity-cost function \cite{blahutComputationChannelCapacity1972} characterises the trade-off between the achievable inverse of the cost-constrained channel capacity (symbols per bit), \(C\), and the maximum tolerable cost (or expense), \(\beta\).
We can characterise the capacity of the storage channel for a given delay by finding the capacity achieving input distribution, \(P_R^*\), which maximises the mutual information between \(\hat{R}\) and \(R\), and thus, achieves the capacity of the channel. Thus:

\begin{align}
    C(d) = \max_{P(R)} I_{P(\hat{R}|R, D)}(R;\hat{R}).
\end{align}

Note that, usually, the set to which the capacity achieving input distribution belongs, is unconstrained. If, however, we have an average power (energy) budget for programming, we must modify the optimisation to obtain the energy-constrained capacity, denoted by \(C_E\), by examining the maximisation problem over the set of valid input distributions - distributions which satisfy the energy budget condition, for a budget \(B\), given as:

\begin{align}
    \mathbb{E}[\mathcal{E}(R)] \leq B
    \label{eq:energy_budget_condition}
\end{align}
where \(\mathcal{E}(\cdot)\) is the energy equation, mapping a given input codeword (resistance, or state) \(\hat{R}=\hat{r}\) to an energy cost.

We can denote the set of input distributions for which the condition \(B\) is satisfied as \(\Gamma_B\), thus the energy-constrained capacity (for the reduced set of possible input distributions) is given as follows:

\begin{align}
    C(d) = \max_{P(R)  \in \Gamma_B} I_{P(\hat{R}|R, D)}(R;\hat{R})
    \label{eq:energy_constrained_capacity}
\end{align}

\section{Data Acquisition and Processing}

\begin{figure}[!htbp]
    \centering
    \begin{tikzpicture}[xscale = 0.55] 
    \draw[->] (-0.2,0.5) -- (4.5,0.5) node[below] {Time};
    \draw[->] (0,-0.5) -- (0,1.5) node[left] {Voltage};
    \draw[thick, red] (0,0.5) -- (0.5,0.8) -- (1.5,0.2) -- (2.5,0.8) -- (3.5,0.2) -- (4,0.5);
    \draw[<->] (0.5,0.5) -- (0.5,0.8);
    \node[above] at (0.8,0.75) {$A_{read}$};
    \draw[<->] (1,-0.2) -- (3,-0.2);
    \node[below] at (2,-0.2) {$T_{read}$};
    \node[above] at (2,2) {READ};
    \node[below] at (2,-0.8) {(a)};
    
    \draw[->] (5,0) -- (7.2,0) node[below] {Time};
    \draw[->] (5.3,-0.5) -- (5.3,2.5) node[left] {Voltage};
    \draw[thick, red] (5.5,0) -- (5.6,2) -- (6.4,2) -- (6.5,0);
    \draw[<->] (6.7,0) -- (6.7,2);
    \node[right] at (6.5,1) {$A_{set}$};
    \draw[<->] (5.5,-0.2) -- (6.5,-0.2);
    \node[below] at (6,-0.2) {$T_{set}$};
    \node[above] at (6,2) {SET};
    \node[below] at (6,-0.8) {(b)};
    
    \draw[->] (8,1.5) -- (11.7,1.5) node[below] {Time};
    \draw[->] (8.3,-0.5) -- (8.3,2.5) node[left] {Voltage};
    \draw[thick, red] (8.5,1.5) -- (8.6,0) -- (10.4,0) -- (10.5,1.5);
    \draw[<->] (10.7,0) -- (10.7,1.5);
    \node[right] at (10.5,0.75) {$A_{reset}$};
    \draw[<->] (8.5,-0.2) -- (10.5,-0.2);
    \node[below] at (9.5,-0.2) {$T_{reset}$};
    \node[above] at (9.5,2) {RESET};
    \node[below] at (9.5,-0.8) {(c)};
    
    \begin{scope}[shift={(0,-4.3)}]
        \draw[->] (-0.2,0.5) -- (12,0.5) node[below] {Time};
        \draw[->] (0,-1.5) -- (0,1.5) node[left] {Voltage};
    
        \draw[thick, red] (0,0.5) -- (0.1,-1) -- (1.9,-1) -- (2,0.5);
        \draw[thick, red] (2,0.5) -- (2.5,0.8) -- (3.5,0.2) -- (4.5,0.8) -- (5.5,0.2) -- (6,0.5);
        \draw[thick, red] (6,0.5) -- (6.1,2.5) -- (6.9,2.5) -- (7,0.5);
        \draw[thick, red] (7,0.5) -- (7.5,0.8) -- (8.5,0.2) -- (9.5,0.8) -- (10.5,0.2) -- (11,0.5);
        
        \draw[<->] (-0.2,-1) -- (-0.2,0.5);
        \node[left] at (-0.1,-0.25) {$A_{reset}$};
        \draw[<->] (0,-0.2) -- (2,-0.2);
        \node[below] at (1,-0.2) {$T_{reset}$};
        \node[above] at (1,2.5) {RESET};
        
        \draw[<->] (2.5,0.5) -- (2.5,0.8);
        \node[above] at (2.8,0.75) {$A_{read}$};
        \draw[<->] (3,-0.2) -- (5,-0.2);
        \node[below] at (4,-0.2) {$T_{read}$};
        \node[above] at (4,2.5) {READ};
        
        \draw[<->] (5.8,0.5) -- (5.8,2.5);
        \node[left] at (5.9,1.5) {$A_{set}$};
        \draw[<->] (6,-0.2) -- (7,-0.2);
        \node[below] at (6.5,-0.2) {$T_{set}$};
        \node[above] at (6.5,2.5) {SET};
        
        \draw[<->] (7.5,0.5) -- (7.5,0.8);
        \node[above] at (7.8,0.75) {$A_{read}$};
        \draw[<->] (8,-0.2) -- (10,-0.2);
        \node[below] at (9,-0.2) {$T_{read}$};
        \node[above] at (9,2.5) {READ};
        \node[below] at (5.5,-1) {(d)};
    \end{scope}
\end{tikzpicture}
    \caption{READ (a), SET (b), and RESET (c) waveforms, and composite RESET-READ-SET-READ cycle  (d).
    \(T_{x}\) and \(A_{x}\) denote waveform periods/amplitudes, respectively.}
    \label{fig:waveform}
\end{figure}
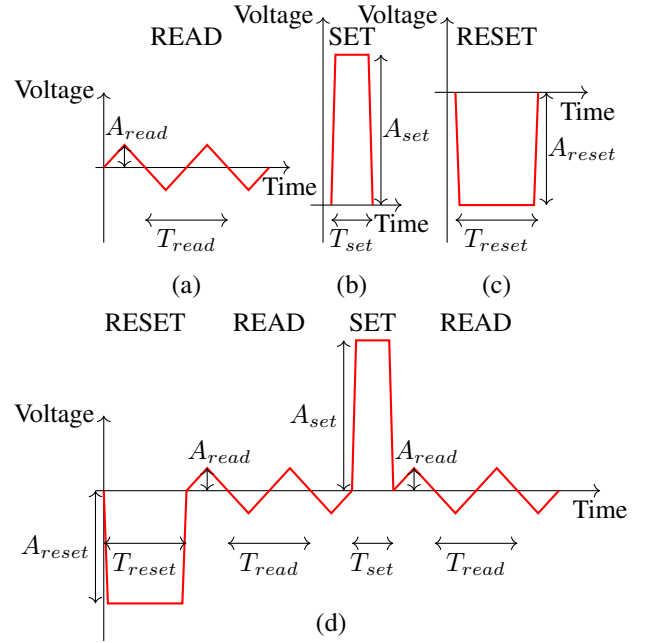

We perform experiments on the Knowm W+SDC memristor - a tungsten doped device, with a thin-layer structure, including a \(Ge2Se3\) active layer~\cite{campbellSelfdirectedChannelMemristor2017}. We use a single memristor in series with a \(100k\Omega\) resistor, \(R_{s}\). \(V_{t}\) is the voltage measured across the entire circuit, while \(V_{s}\) is the voltage across the series resistor. From these measurements, we can determine the voltage across and current through the memristor using \(V_t-V_s\) and \(V_s/R_s\), respectively.
We use square pulses with rising and falling edges to program (change state) and small-amplitude triangular waves to read, as shown in Figure~\ref{fig:waveform}.

\subsection{Signal Alignment for Unknown Systematic Offset Removal}
\label{sec:data_preprocessing}

\begin{figure}[!htbp]
    \centering
    \includegraphics[width=\linewidth]{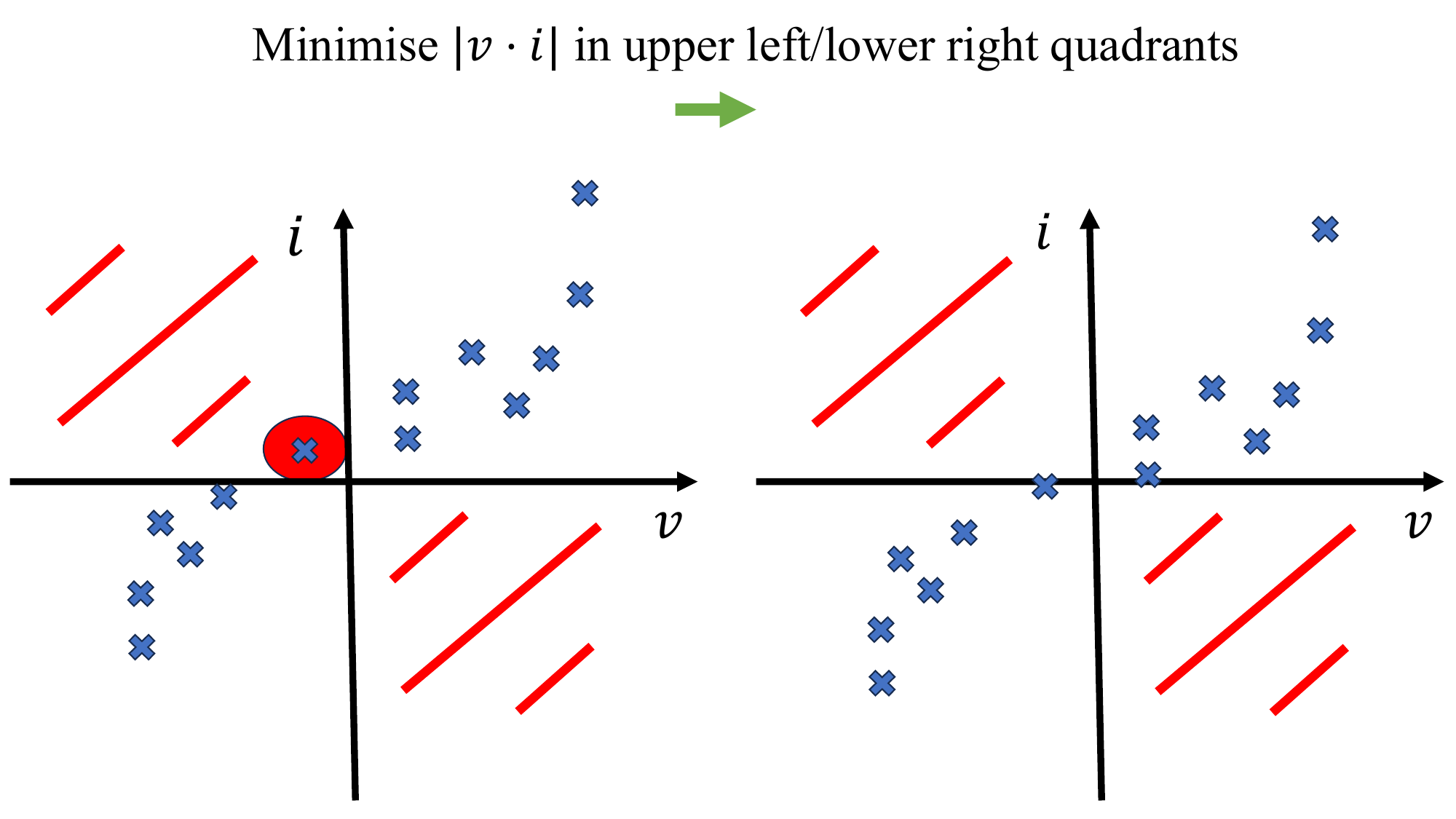}
    \caption{To reverse unknown systematic offsets, we minimise \(|v \cdot i|\) in quadrants that should be empty for passive devices.}
    \label{fig:offset_alignment}
\end{figure}

To correct for timing misalignments (along the time axis i.e. x axis), we used a simple minimum absolute-value method to detect and discard \(N_{\text{discard}}\) initial points. We then trimmed the trailing points to retain \(N_{\text{period}}\) datapoints per measurement.
Next, we applied automated offset correction (along the signal axis i.e. y axis) through an iterative optimisation procedure that considered the current and voltage jointly by minimising the magnitude of \(|v \cdot i|\) in the upper-left and lower-right VI quadrants, under the assumption that the current and voltage for the aligned signal should be of the same sign for passive electronic devices. This is illustrated in Figure~\ref{fig:offset_alignment}.

\subsection{Memristive State Estimation}
\label{sec:state_estimation}

In our experiments, to estimate the memristive state, we adopt the state estimation procedure proposed in \cite{hajtoStateCharacterisationSelfDirected2025}, 
weighting VI pairs according to their uncertainty to produce a minimum variance estimate. The factors for a tuple of measurements \((v_n, i_n)\) - denoted \(c_n\) - are given as:

\begin{align}
    c_n &= \frac{K_n}{\sum_j K_j},\\
    K_n &= \left(\frac{1}{G_m v_n + \alpha_1 (e^{\beta_1 \cdot v_n} - 1) + \alpha_2 (1 - e^{-\beta_2 \cdot v_n})}\right)^{-2},
    \nonumber
\end{align}
and the minimum variance state estimate, \(R_{\text{est}}\), is given by weighting the individual estimates derived from respective pairs:

\begin{align}
    R_{\text{est}} = \sum_n \frac{c_n \cdot i_n}{G_mv_n + \alpha_1 (e^{\beta_1 \cdot v_n} - 1) + \alpha_2 (1-e^{-\beta_2 \cdot v_n})},
\end{align}
which is simply Equation~(\ref{eq:vi}) rearranged, scaled by the factors that account for differences in the variance of the estimates derived from measurements at different magnitudes.

The state variable approximately corresponds to a \textit{scaled} conductance over low voltage input magnitudes. Throughout the paper, we instead use the inverse of the state variable \(x\), which corresponds to a \textit{scaled} approximate resistance valid for low magnitude voltage inputs (scaled approximately by the magnitude of the parameter \(G_m\)), and we thus give it dimensions \(\Omega\).

\section{Energy Cost Function Modelling}
\label{sec:energy_characterisation}

A generic logarithmic trade-off was noted between the energy expenditure and the programmed resistance state in \cite{el-geresyEnergyConstrainedInformationStorage2024}. Although a general form of the energy cost function was characterised, the exact parameters of this function will be device dependent, and experimental evidence is needed to determine them.
Here, we perform experiments that aim to characterise the nature of the energy cost function, \(\mathcal{E}\), for \ac{SDC} devices.

We define the energy cost function, \(\mathcal{E}(R)\)
as follows, being the cost associated with bringing the device from an equilibrium (resting) state \(R_{\text{eq}}\), to a given target state \(R\):

\begin{align}
    \mathcal{E}(R) = \left| A \cdot \ln{\left| \frac{B}{R} \right|} \right|,
    \label{eq:energy_cost_function_general_form}
\end{align}
where \(A\) and \(B\) are the fitting parameters, with \(A \triangleq \tau_{\text{final}} \cdot K^2 \cdot \left( \frac{1}{R_{\text{final}}} - \frac{1}{R_{\text{eq}}} \right)\) and \(B \triangleq R_{\text{eq}}\), with \(R_{\text{final}}\) and \(\tau_{\text{final}}\) being the minimum achievable resistance and the duration of a programming pulse needed to bring the device from \(R_{\text{eq}}\) to this resistance.

\subsection{Energy Experiment}
\label{sec:state_dataset}

We conducted a series of experiments aimed to determine the energy required to program the memristor into various resistive states to determine the parameters of the energy cost function.
Each experiment consists of a series of 100 cycles of a signal, where one period of the signal is composed of: a \(T_{\text{reset}}=3\text{ms}\) RESET, 2 periods of a \(T_{\text{read}}=1\text{ms}\) READ, a \(T_{\text{set}}=2\text{ms}\) SET, and finally 3 periods of a \(T_{\text{read}}=1\text{ms}\) READ signal as shown in Figure~\ref{fig:waveform}(d). In a single signal period, the device is therefore reset to the equilibrium state, measured, then set to a given high conductance state and measured again. The process was repeated for 12 different SET pulse amplitudes uniformly spaced in the range \(A_{\text{set}} \in [500mV, 1600mV]\), allowing us to compare the state reached in response to pulses of different pulse energies.
For each SET amplitude, four repeated experiments were performed. RESET pulses were scaled proportionally to the SET amplitude for each experiment, since a deeper state would be reached.
Following data collection, we filtered the collected signals to remove several with erroneous data.

\subsection{Cost Function Parameter Estimate}
\label{sec:energy_cost_function_parameter_estimate}

We make use of the data collected to estimate the parameters \(A\) and \(B\) of the energy cost function in Equation~(\ref{eq:energy_cost_function_general_form}) according to a least square parameter optimisation procedure.
The model fitting procedure resulted in best parameters of \(A = 0.000239506\) and \(B = 7.1853 \times 10^6 \Omega\).
The value of \(B\) is the estimated equilibrium state, which is in agreement with the known order of magnitude of approximately \(10 \times 10^6 \Omega\).

\begin{figure}[!htbp]
    \centering
    \includegraphics[width=\linewidth]{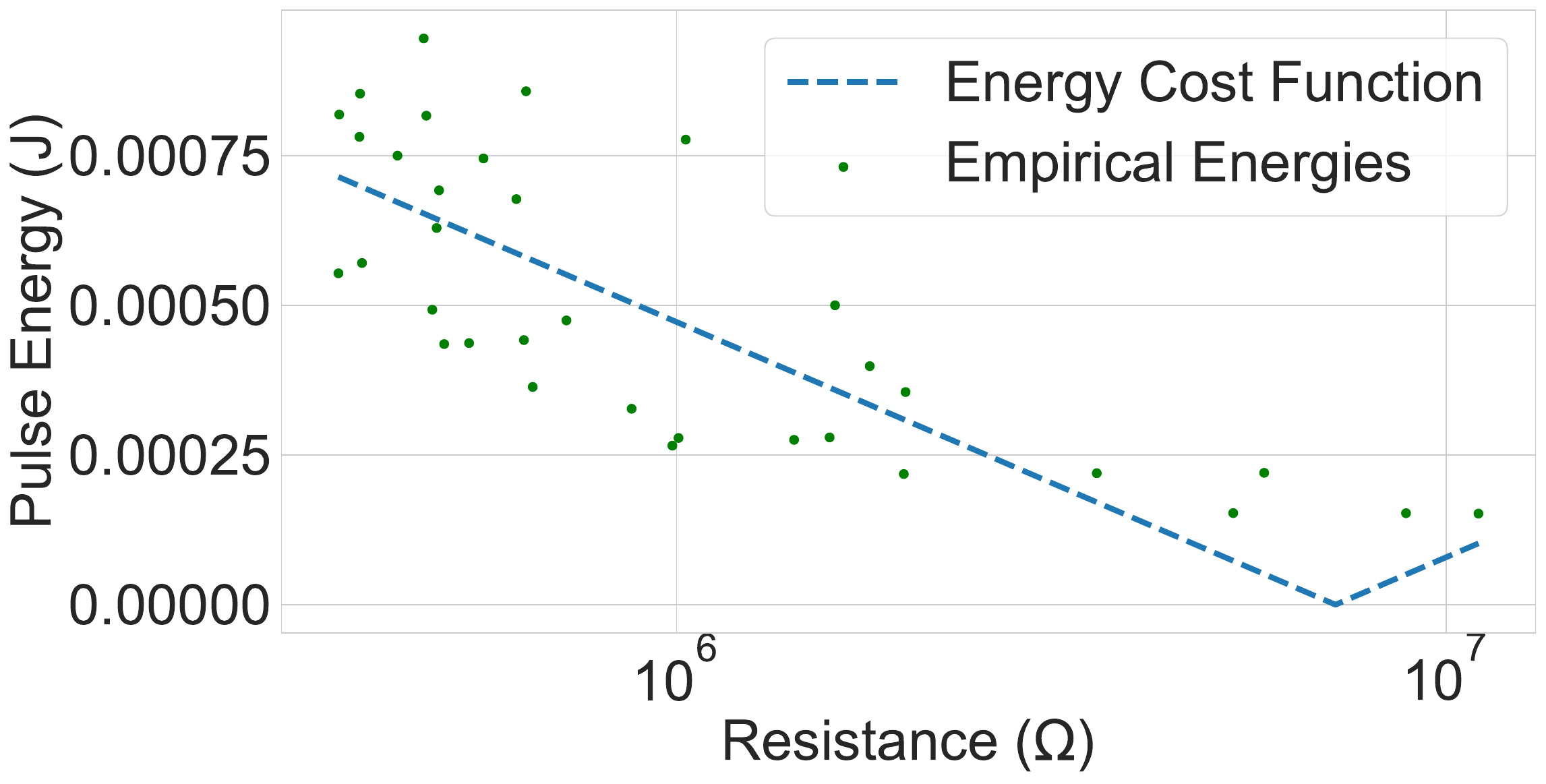}
    \caption{The estimated empirical energies of the SET pulses for experiments with set pulse magnitudes in the range \([500mV, 1600mV]\), plotted alongside the energy estimated for bringing the device into the given state from the equilibrium (high resistance) state according to the proposed energy cost function.
    Programming noise results in a distribution of samples around the predicted cost function \cite{zhengErrorResilientAnalogImage2018}.
    }
    \label{fig:energy_cost}
\end{figure}

\section{Delay Conditional Distribution Modelling using cGAN}
\label{sec:state_stability}

In this section, we perform retention experiments to characterise 
the stability of device states, with a view to estimating the continuous conditional distribution \(P(\hat{R}|R,D)\) using the \ac{cGAN} model that is trained on the retention data.

\subsection{Retention Experiment}
\label{sec:retention_experiments}

\begin{figure}[!htbp]
    \centering
    \includegraphics[width=\linewidth]{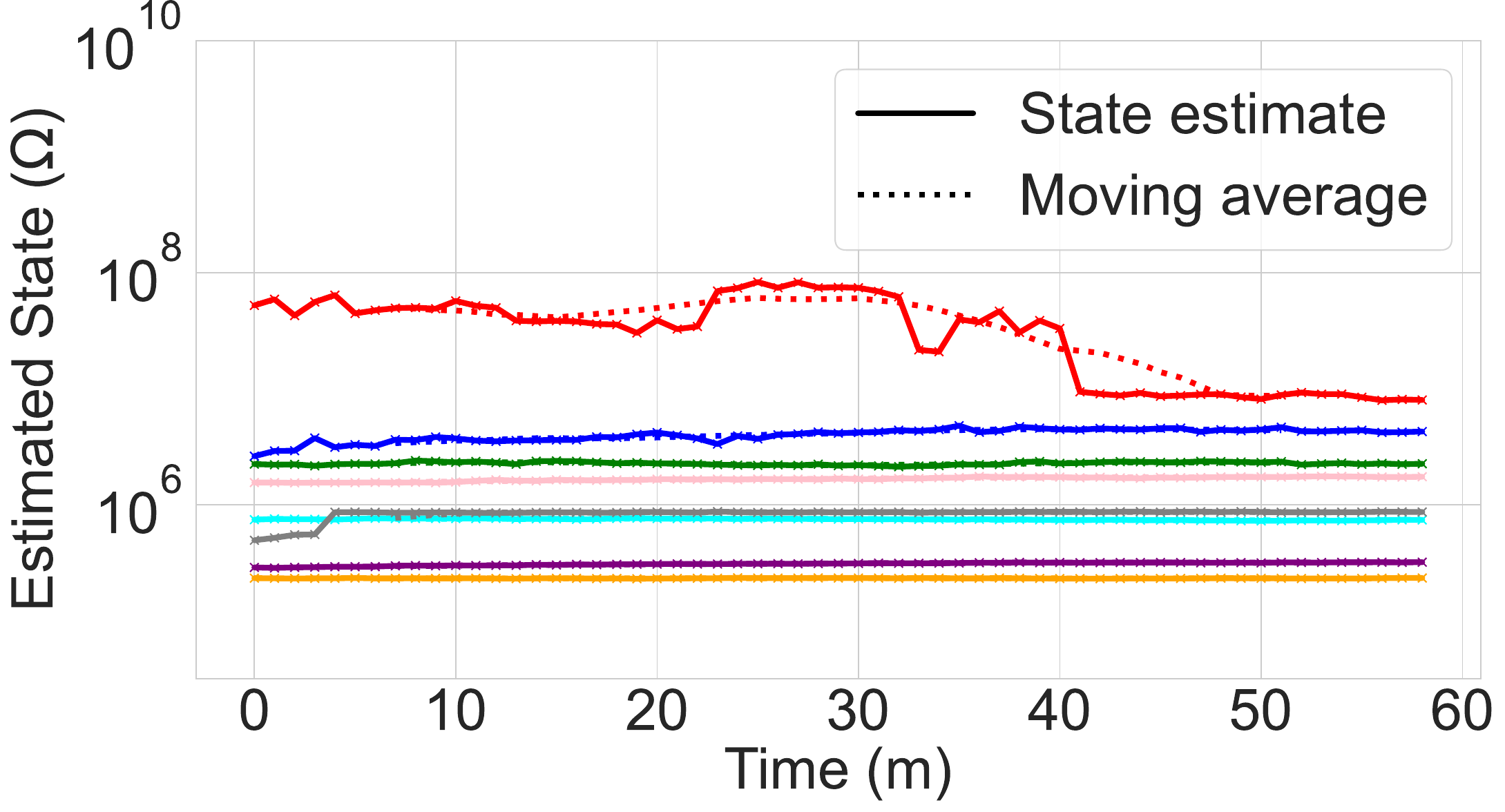}
    \caption{State characterisation of the memristive state for retention experiments with different (arbitrary) initial states. Dotted lines are moving averages, while solid lines are the estimated states. The estimated equilibrium state (B) is $7.1853 \times 10^6 \Omega$, with states expected to tend toward this value over time. For the red series with the highest starting state, the state trends down as the device starts above equilibrium.}
    \label{fig:retention_states}
\end{figure}

We conducted 9 experiments, at the start of each of which the memristor was programmed to an arbitrary initial state, and then allowed to evolve over time.
We measured the device states over the course of 60 minutes (1 hour). We applied the state estimation procedure described in Section~\ref{sec:state_estimation} to estimate the state every minute from READ waveform measurements. Figure~\ref{fig:retention_states} shows the results of the state characterisation applied for the different time series.

\subsection{Modelling Resistive Drift}

\begin{figure}[!htbp]
    \centering
    \includegraphics[width=\linewidth]{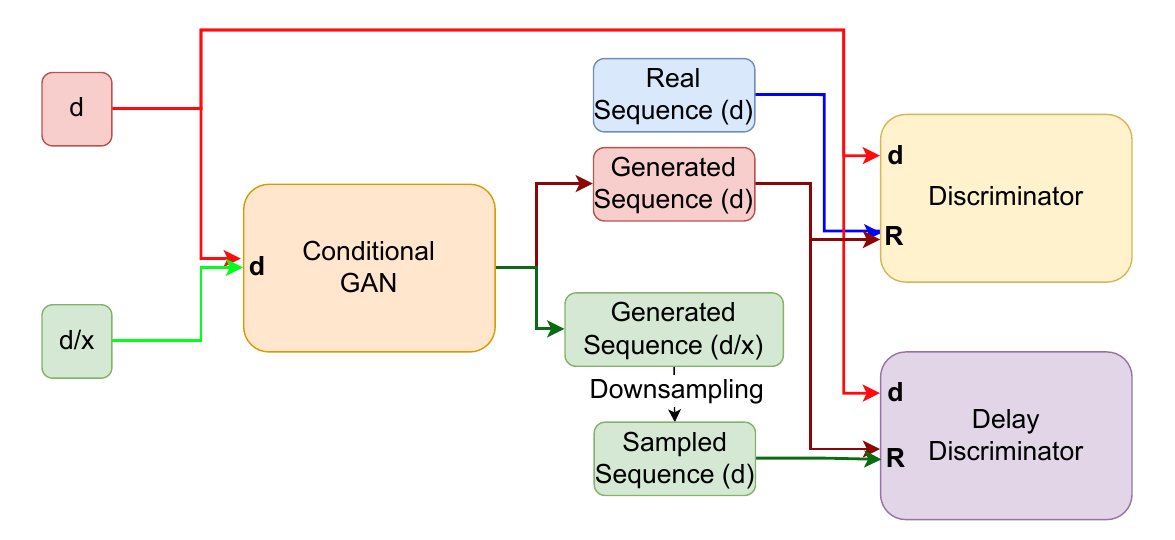}
    \caption{The training scheme for the \ac{cGAN} including the delay discriminator. Generated sequences are produced recursively.}
    \label{fig:conditional_gan}
\end{figure}

\begin{figure*}[!htbp]
    \centering
    \subfloat[]{\includegraphics[width=0.33\linewidth]{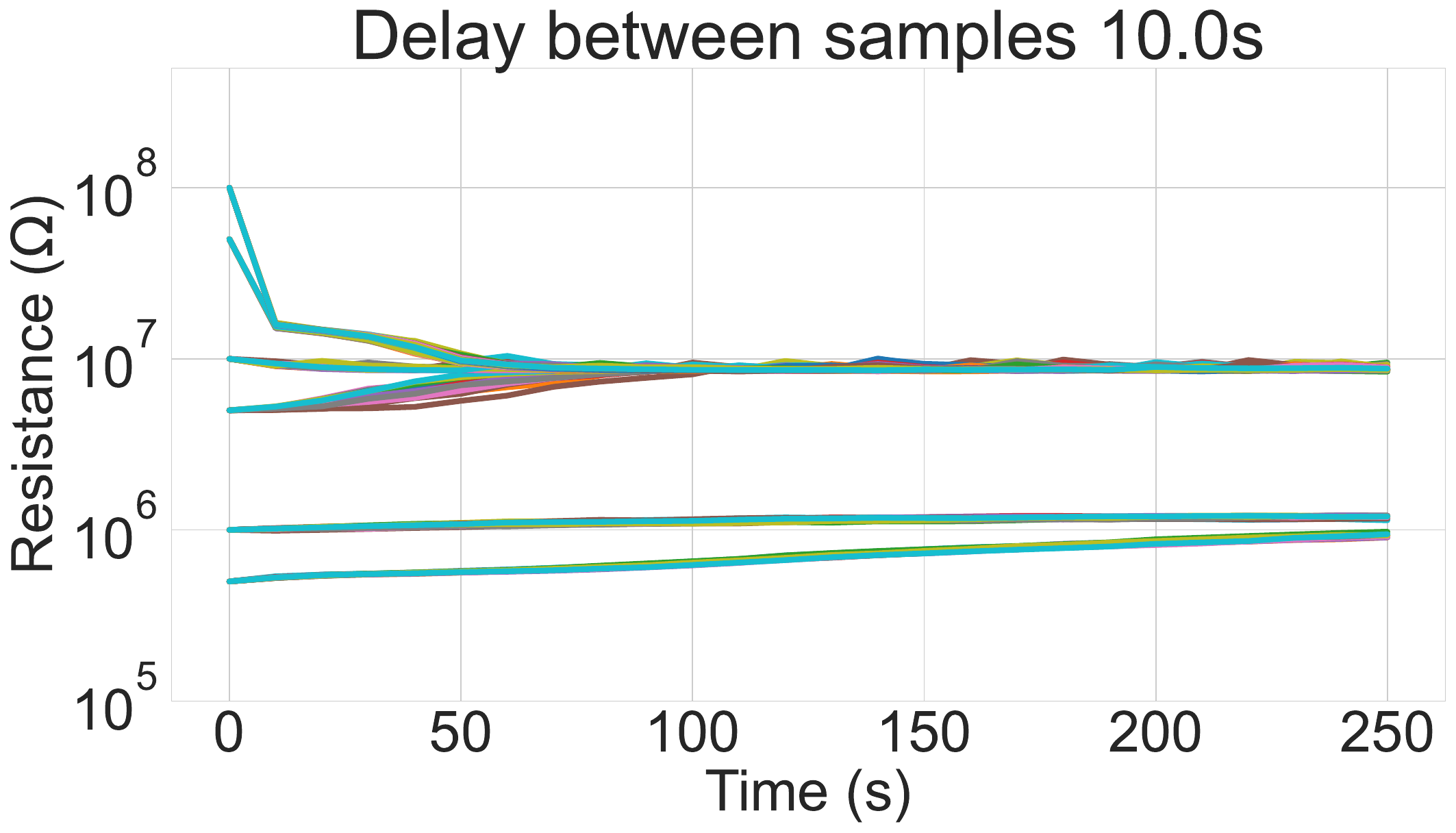}}
    \hfill
    \subfloat[]{\includegraphics[width=0.33\linewidth]{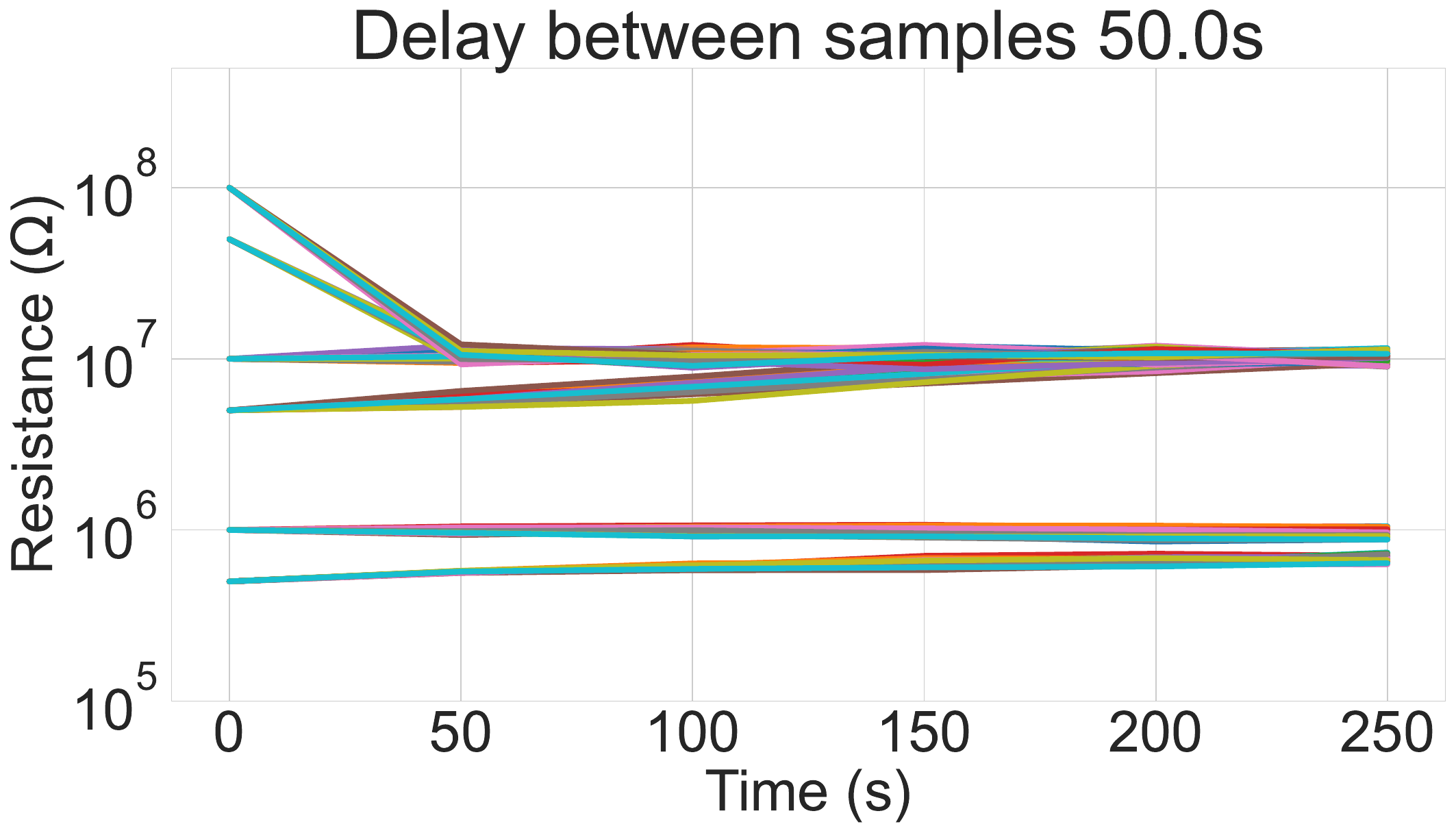}}
    \hfill
    \subfloat[]{\includegraphics[width=0.33\linewidth]{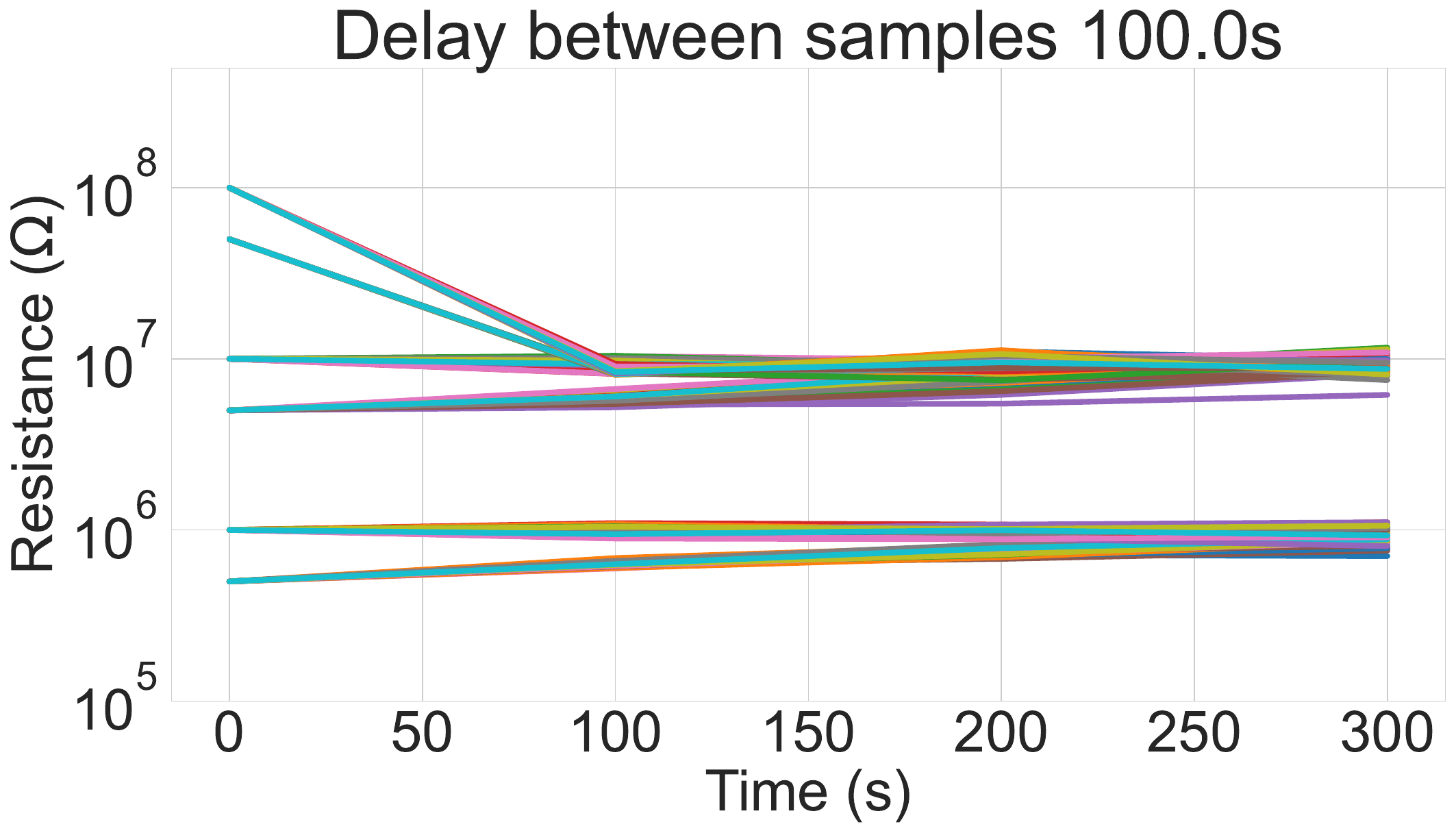}}
    \caption{Series generated by the \ac{cGAN}, showing consistency in the generation of resistive drift data, linearly interpolated based on the generated points. Each coloured line represents a different generated series for starting resistances in \([5\times10^5, 1 \times 10^8]\).}
    \label{fig:cgan_training_results}
\end{figure*}

We use a \ac{cGAN} generative modelling approach to model the conditional resistive drift distribution based on the drift data, with an approach involving two discriminators in order to produce a delay-consistent model as described in \cite{el-geresyDelayConditionedGenerative2024}.
The advantages of a generative modelling approach include the final model's data-driven nature and end-to-end differentiability \cite{yeDeepLearningBasedEndtoEnd2020}. The additional delay discriminator of the \ac{cGAN} allows us to extrapolate beyond the range of the original datapoints with delay consistency. This technique is especially useful given the relatively small number of datapoints. Our model is trained on the 9 retention time series, each of length 59 minutes, with samples every minute. A primary discriminator discriminates between sequences of resistances from the dataset, sampled at a given delay, and generated sequences for that delay. The auxiliary ``delay'' discriminator is trained to distinguish between generated sequences sampled at different timescales.
We use the model and parameters proposed in \cite{el-geresyDelayConditionedGenerative2024}, changing the length of training sequences to be 4 and the maximum sampled training delay for the main discriminator to be 10s (with a total sequence span of 40s therefore less than the maximum of 60s). Consequently, we adjusted the maximum delay discriminator training delay to 200s. The total generator loss is composed of two summed binary cross-entropy losses for each discriminator. This is shown in Figure~\ref{fig:conditional_gan}.
 Figure~\ref{fig:cgan_training_results} shows generated series for a number of delays, alongside the 9 time series used as training data. The device state tends to an equilibrium value on the order of \(10 M\Omega\), consistent with our estimate of the equilibrium value derived in Section~\ref{sec:energy_characterisation}.

\section{Energy-Information Trade-Off}
\label{sec:blahut}

Having determined the nature of the energy cost function in Section~\ref{sec:energy_characterisation}, as well as a continuous model of the conditional drift distribution in Section~\ref{sec:state_stability}, we can jointly consider both results to characterise the energy-information trade-off.

To determine the channel capacity of the devices for different power budgets, we can make use of the cost-constrained Blahut-Arimoto algorithm \cite{blahutComputationChannelCapacity1972}. This algorithm allows for the estimation of the capacity of a discrete communication channel with cost constraints on the input symbols, yielding an estimate of the capacity accompanied by the proposed capacity achieving input distribution for a given cost. The cost is not specified directly, but rather parameterised by a value \(s \in [0, \infty]\), yielding a family of capacities for a variety of cost constraints, each corresponding to a given \(s\).

\subsection{Quantisation}

To produce an estimate of the capacity, the Blahut-Arimoto algorithm requires the use of a discrete set of symbols. As such, we quantise the resistive states of the memristor into a choice of \(q\) bins for the purposes of estimation. We use \(q = 100\) equally spaced bins in the (inclusive) interval \([100k\Omega, 20M\Omega]\).
We use the \ac{cGAN} model of \(P(\hat{R}|R, D)\), marginalised over different delays \(D=d\), to compute a family of delay conditioned distributions. For each distribution, we quantise the input resistances uniformly in the range \([100 k\Omega, 1 M \Omega]\), split over \(q=100\) quantisation bins, with the centre of each bin being assigned as the quantisation level.
We estimate the discrete conditional distribution (according to the chosen bins) for each value of \(R_{\text{init}}\) for a given delay through a statistical estimate obtained through computing the relative frequencies of the values of \(R_{\text{final}}\) for \(n=1000\) simulations per resistance value. We compute discrete conditional distributions for delays in the set \(\{10, 50, 100\}\).

In order to calculate the (energy) cost of symbols for use in the cost-constrained algorithm, we make use of the estimated energy cost function, \(\mathcal{E}(r)\) for the discrete centroid values of \(R_{\text{init}}\) as shown in Equation~(\ref{eq:energy_cost_function_general_form}), with parameter values for \(A\) and \(B\) as given in Section~\ref{sec:energy_cost_function_parameter_estimate}.

\subsection{Results and Discussion}

\begin{figure}[!htbp]
    \centering
    \includegraphics[width=\linewidth]{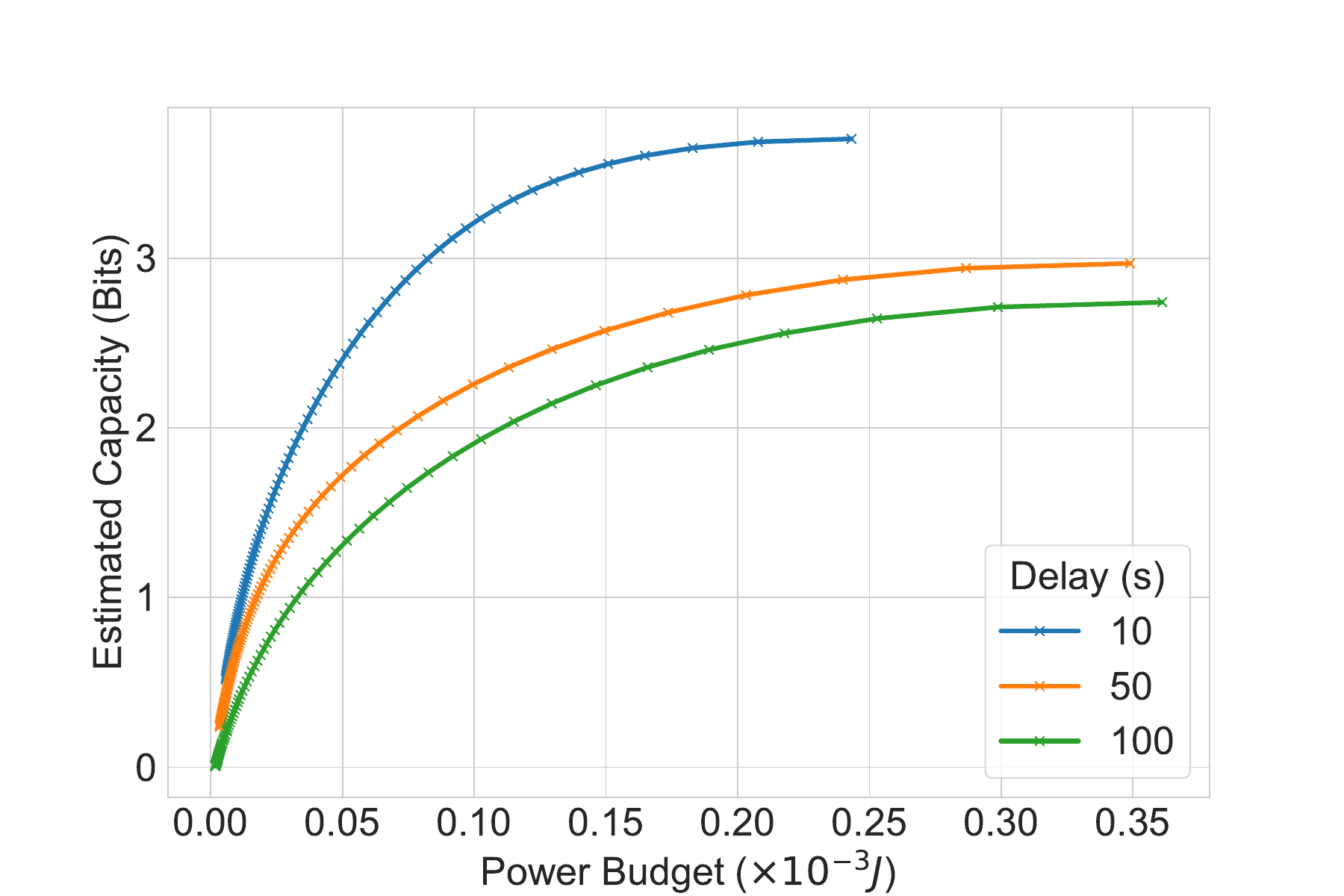}
    \caption{The cost-capacity curves for 3 different delays between writing and recovery on the memristor, showing the estimated trade-off between average energy per symbol (power consumption) and the achievable information capacity.}
    \label{fig:cost_capacity}
\end{figure}

We estimate the discrete conditional distribution for each delay in the set, and then calculate energy-capacity pairs across 100 equally spaced values of the parameter \(s\) in the range \([10\times10^{-10}, 10\times10^{4}]\) to plot the cost-capacity curves.
The results of these experiments are shown in Figure~\ref{fig:cost_capacity}.
We see that not only does the maximum asymptotic capacity of the device change for different delays, with a higher asymptotic capacity observed for smaller delays (a lesser degree of loss of information), but we additionally observe that the energy required to achieve this capacity also changes. In the case of smaller delays (e.g. 10m), a smaller amount of energy can be used to come close to the higher asymptotic information capacity, whereas in the case of larger delays (50m and 100m), we see that more energy is required to achieve the (successively lower) asymptotic capacities.

Assuming a near capacity-achieving code is used to store for a certain target delay and energy, we will observe a threshold behaviour, where the information will be recovered up until this target delay, while the error probability will drastically increase beyond this. One way to achieve more graceful data storage as a function of delay is to consider a superposition approach, where multiple data streams, each targeting a different delay, can be superposed on the device, akin to a strategy for communication over a Gaussian fading channel \cite{shamaiBroadcastStrategyGaussian1997}.

\section{Conclusions and future work}

In this work, we have investigated the use of one particular kind of memristive device for storage, characterising the nature of the energy-information trade-off for the device. We began by characterising the uniquely identifiable states in the device, according to mathematical modelling inspired by a proposed physical model of the device. We then investigated the energy expenditure required to bring the device into a range of lower resistive states from an initial high resistive state, using square voltage pulses of a fixed duration and varying amplitude. Finally, we used this observed trade-off to characterise the energy-information trade-off for an exemplary binary storage scheme on the devices, calculating the energy-constrained device capacity for a range of energy budgets and highlighting the nature of the energy trade-off.

\begin{acronym}[]
    \acro{CBRAM}{Conductive Bridge RAM}
    \acro{cGAN}{conditional Generative Adversarial Network}

    \acro{ECM}{Electrochemical Metallization}

    \acro{Generalised MSS}{Generalised Metastable Switch Model}

    \acro{JSCC}{Joint Source-Channel Coding}

    \acro{MRE}{mean relative error}
    \acro{MRSE}{mean relative squared error}
    \acro{MSE}{mean squared error}
    \acro{MAE}{mean absolute error}

    \acro{PCM}{Phase Change Memory}

    \acro{SDC}{Self-Directed Channel}
    \acro{SNR}{Signal-to-noise-ratio}

    \acro{TCM}{Thermochemical Mechanism}

    \acro{VCM}{Valence Change Mechanism}
\end{acronym}

\bibliographystyle{IEEEbib}
\bibliography{references}

\end{document}